# HUMAN-DATA INTERACTION IN HEALTHCARE

Federico Cabitza and Angela Locoro

*Dipartimento di Informatica, Sistemistica e Comunicazione, Università degli Studi di Milano-Bicocca*
*Viale Sarca 336 20126, Milano, Italy (Phone: +39-02-64487888); E-mail: {cabitza,angela.locoro}@disco.unimib.it*

**ABSTRACT**

In this paper, we focus on an emerging strand of IT-oriented research, namely Human-Data Interaction (HDI) and on how this can be applied to healthcare. HDI regards both how humans create and use data by means of interactive systems, which can both assist and constrain them, as well as to passively collect and proactively generate data. Healthcare is a challenging arena to test the potential of HDI towards a new, user-centered perspective on how to support and assess data work, especially in current times where data are becoming increasingly big and many tools are available for the lay people, including doctors and nurses, to interact with health-related data. This paper is a contribution in the direction of considering healthcare data through the lens of HDI, and of framing data visualization tools in this strand of research, in order to let the subtler peculiarities among different kind of data and of their use emerge and be addressed accordingly. Our point is that doing so can promote the design of more usable tools that can support data work from a user-centered and data quality perspective.

## 1. INTRODUCTION

Twenty-five years ago, medical informatics was defined as "dealing with the storage, retrieval and optimal use of biomedical data" (Shortliffe et al. 1990). At that time, little emphasis was put on the practices of data production, that is on how medical practice, and single stories of illness, care and recovery are represented, accounted and "datafied" in some objective manner. However, these practices, which include policies, rules, habits, conventions, tools and techniques, have always been intertwined with and affected by the available ITs, as well as by the expectations of the stakeholders on how to make sense and use of health-related data. Different perspectives on these expectations, and on what valuable health data are, lead to manifest chasms between primary use and other uses of health information, as often discussed in the specialist literature (Fitzpatrick and Ellingsen, 2013). To try to cross these chasms, we need to create the suitable language to describe the differences and give some operational definitions.

We distinguish between three different macro-types of data and the related processes in which these data are either produced, processed or consumed: namely, *primary data*, which come from a broad range of sources and are produced *within* a caring process to make its unfolding seamless and smooth (Berg, 1999); *secondary* data that are derived from the primary data for purposes different than care, like accounting and medical billing (Abdelhak, Grostick, & Hanken, 2012); and *tertiary* data, that are produced from the secondary data for any unanticipated need of the potential consumers of health services (see Figure 1).

To illustrate this tripartition, an analogy from the agriculture domain can be drawn (Locoro, 2016): primary data are like the produce of the land, which farmers grow for themselves as well as the external market. Secondary data are the product of a transformation of these primary data, like the one performed in food industry where vegetables are cleansed and chopped. Tertiary data are further transformed from secondary data to make them more easily consumable, that is suitable for and conveyed to a broader population of consumers in terms of information services, like fresh-cut vegetable products can be seen as the service to have vegetables already ready-to-eat.

The definitions mentioned above shed light on the relationship between data and their uses, which cannot be overstated. The tripartition that we propose reflects the different uses and practices in which data are produced and consumed and it calls for a specific area of research focusing on how to support people in interacting with their data of concern and in gaining insight from them: Human-Data Interaction (HDI). HDI is both a phenomenon and a research field focusing on this phenomenon. As a phenomenon that is object of research it is the kind of action where, on the one hand data are produced, processed and exploited by humans; this encompasses, at the two extremes of the action spectrum, both the *datification* of facts, that is the process by which portions of the reality of interest are translated into the domain of words, symbols and numbers (through coding, classification, and measure[1]); and what we call *data telling*, that is the creation of accounts and stories that human can tell according to the data they make (a) sense of. On the other hand, HDI also regards actions in which humans are affected by data, that is in which their interpretations, opinions, beliefs and decisions are made by, or better yet according to, data[2]. As a research area focusing on this kind of interaction it lies at the intersection of data science, data visualization and human-computer interaction. Researchers involved in HDI *study* how humans record and use data by means of Information Technologies; and *design* interactive systems that allow humans to retrieve and explore complex data sets in order to gain value in their learning, insight in their decisions, and feedback in their action[3]. In what follows we will justify our triparted perspective and propose it as a conceptual framework to focus on the peculiarities and needs of primary and tertiary users, and therefore contribute in the HDI field of research applied to healthcare data.

## 2. DISTINGUISHING WAYS OF INTERACTING WITH HEALTH DATA

For a straight metonymy, we intend primary data as entangled with the primary use of data, by definition; secondary data with secondary use(s); tertiary data with tertiary uses. The "primary use" of health information is "to use it to directly support patient care", both by aiding medical decision-making and by ensuring continuity of care by all providers, that is both interpretation of medical signs (represented by data) for decision making, and coordination among the actors involved "around" the patient (Berg, 1999). Secondary use regards both other uses of the same data collected for the primary use within the administrative domain, and the generation of derivative data for other aims than care, like billing and reimbursement, performance and care quality evaluation, resource planning and management, service design and public policy making. Tertiary use regards the heterogeneous uses through which tertiary data are put to the test of users' life: this can encompass the publication and dissemination of valuable indications for the citizens and the taxpayers about the available healthcare services, so as to orient them in the choice of the best healthcare provider (or just the closest one to their shelters – cf. Cabitza et al. 2015b); the policy makers and facility or agency manager, to facilitate the monitoring of suitable outcome and performance indicators and to enable the benchmarking and comparison of care facilities; as well as the researchers and epidemiologists to make sense of the great numbers of "similar" patients that are treated and detect trends and patterns of treatment appropriateness and efficacy.

On a more general level, three concepts characterize uses and data in each of these three ambits: primary use is related to what has been often called "*data work*" (Bossen et al. 2016). Secondary uses with *data storing and processing*. Tertiary uses with *data value*. Data storing and processing are self-explanatory concepts in computer science and regard the set-up of efficient data bases and the automation of accurate procedures by which to structure data, run searches and extract complex reports for disparate purposes. As secondary data and secondary processes usually entail little interaction by humans and these are always IT specialists and consultants who have always considered usability a relative concern, we will focus on the primary and secondary domains, and hence on data work and data value.

---

[1] To this respect, datification is a process of both dis-closure of the representable reality, as well as the closure of what is (willingly or unaware) left tacit behind. Datification constructs facts, but it can also entail the (irreversible) destruction of context, through the emergence of text out of human experience.

[2] Another way to see HDI is the *micro* level of *data work*, that is the work humans perform both on, with and by data, a co-productive process that also shapes human data-intensive agencies.

[3] The reader should mind that *human-data interaction* is not *human data* interaction, that is interaction with human data. We intend HDI as focused not solely on the interaction of humans with human-related data, or with one' own personal data, but *also* on that kind of data and the related dimensions, which have been recently considered by other scholars (e.g., Crabtree and Mortier, 2015).

Data work is quite never defined in the specialist literature; although it is often mentioned, it is a slippery notion. We see it as a modern (but not at all recent) coinage after the word *paperwork*, of which it represents an abstraction, with respect to the medium of data representation; but also an extension, with respect to what people manage as data of their interest (besides accounting and resource management). As such data work is not only "working *on* data", typically producing new data in accounting for and recording a faithful representation of the work done; but also *that* portion of work that can be accomplished only by relying on some accurate data, i.e., "working *by* data". These two kinds of work are usually so deeply intertwined that distinguishing between them is useless and probably wrong: in the healthcare domain, the studies by Berg (Berg, 1999), for instance, clearly show that clinicians record data on the patient record not only to accumulate data for archival reasons (and for many other secondary uses), but also to coordinate with each other, articulate the resources around a medical case, and take informed decision in a written, distributed communication with themselves and the other colleagues taking care of the same patient. In healthcare, data work regards the additional effort paid by caregivers in making the record a "working record" (Fitzpatrick 2004), that is a resource capable of keeping disparate competencies and roles bound together and connected around the same cases over time and space.

On the other hand, tertiary use of tertiary data regards the concept of *data value*, that is how to exploit data to achieve some relevant goal for the user (even just to be informed about something), so that we can say that the user *finds* value in the data (also: she finds data valuable), or she *gains* value to some aim. Whether value is already in the data or rather it is created by an active stance of its consumers looks like an idle question: in a HDI perspective, value is the *result* of interacting with data and being capable to exploit them by tertiary users, that is lay people moved by unexpectable motives and toward unanticipated aims. In other words, on one hand data have got value if they are true and have been made accessible and comprehensible; on the other hand, their value lies in the comprehension itself, in the acquisition of true information, in learning notions, techniques, practices, and in the resulting knowledgeable behaviors, in their turn producing some positive effect, on either the single person or her community[4]. Thus, data value can be defined as the potential of data (for its intrinsic qualities) to enable processes of value creation by someone for anyone (either herself, or the others) in some given context.

## 3. HDI CONTRIBUTIONS FOR THE QUALITY OF INFORMATION

While this tripartition is general enough to be applied in any domain where many different roles partake to the generation and transformation of data in the same supply chain, so to say, from the practitioner working in the field, to the final consumer of information services, its application to healthcare can be justified for its role in addressing the problem of the still relatively low quality of the data of patient records (Liaw et al. 2013), both in regard to completeness, accuracy and timeliness (Cabitza & Batini, 2016). Indeed it allows, on one hand, to highlight differences, separate concerns, and identifying ambits of use that are irreducible to one another; on the other hand, it allows to see the (existing) "seams" between these ambits (Berg and Goorman 1999) and therefore to address cooperation and interoperability without reckless neglect.

In regard to this latter point, it is time to acknowledge that chasms between primary and other uses exist and they go beyond the usual tension between clinical vs. administrative purposes because it is the concept of "adequate quality" that should be adopted to assess care-oriented data; likewise, crossing the chasm between the secondary and tertiary domains requires more than merely making secondary data more open and accessible, because also the end consumers' readiness to access, comprehend and exploit them, as well as their unanticipatable purposes, are to be considered. Thus, the challenge to bridge the chasms mentioned above can be seen in terms of the problem of reusing primary data, which is "always entangled with the context of its production" (Berg and Goorman, 1999), in different contexts (either secondary or tertiary ones).

---

[4] Obviously we are aware that the value of something, and hence also data, can be appraised differently than by adopting a consumer-oriented perspective, which regards what in classical political economy is denoted as "use value". Value can be a function of the resources involved or consumed to get the data (mainly by primary users), to verify and cleanse them, and make them fit to interoperable formats typicaly in secondary processes); or, simply, a function of how much someone is going to pay to obtain them (cf. the idea of economic value, that is so common in this age of splendour of online advertising). These different perspectives reflect different conceptions of the idea of value, as it has been formulated in centuries of economic and theoretical speculation (see (Dobb, 1975) for a classic review of those conceptions).

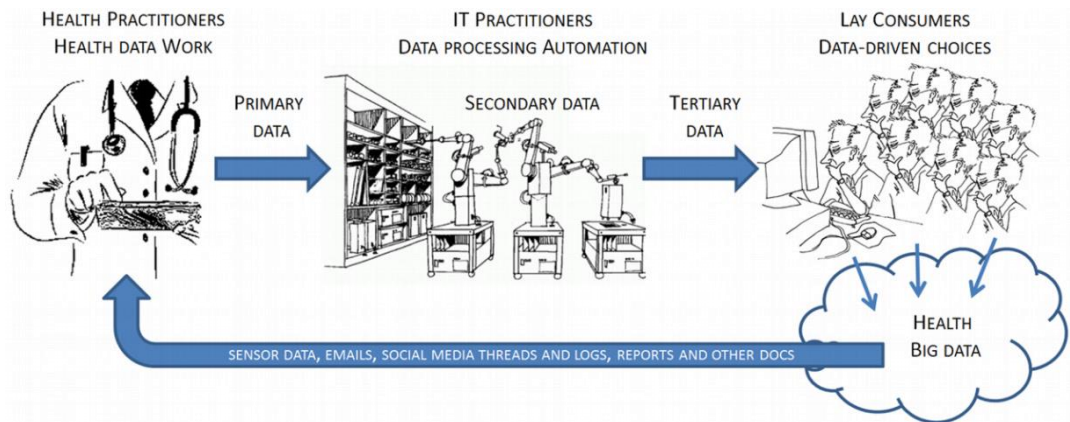

Figure 1 The three kinds of data and ambits proposed for HDI in healthcare.

As Berg and Goorman note, reuse is possible only if data are made "transportable", that is sufficiently disentangled from their context of production; and this can occur only if specialized *additional* work is performed on data. So it is not a matter of improving, automating or changing current data work, but rather to invest on new and different data work, and corresponding new organizational roles (e.g., data nurses).

In regard to the separation of concerns, our distinction between primary, secondary and tertiary data is quite different from other frameworks discussed in the Data Quality literature: for example, the one inspired by the manufacturing domain (Shankaranarayan et al., 2003) which distinguishes between *raw data*, *component data items*, that is semi-processed information and *information products*, which are composed out of these items. Primary data are not necessarily raw (Gitelman, 2013), because they are meaningful to and usable by the primary consumers that generated them. On the other hand, secondary and tertiary data are unfinished information products: the former ones are resources for specialist work (both clerical and managerial) within specific organizational boundaries and processes, while tertiary data result from the enactment of information services conceived for the non specialist and the external consumers, that is for the public. The distinction introduced by Shankaranarayan and his colleagues focuses on an incremental definition of the information product, and is often mentioned in regard to how Data Quality (DQ) can be assessed, monitored and continuously improved (e.g., Wang, 1998)

Differently from the Information Product perspective, our triparted perspective focuses on the different roles that produce and consume different information products that can be considered "definitive" only in relation to their context of use. Consequently, a HDI approach to DQ issues grounds on the interpretation of quality of data as their "fitness for use" (Wang & Strong, 1996). Our proposal is aimed at detecting and taking the differences from the various uses in healthcare seriously (Bossen et al. 2016).

According to our metaphor, secondary data and secondary data-centered processes are like industrial processes, that need standardization to achieve regularity and efficiency and that involve relatively few people with specific skills and motives, like the competent and trained operators of assembly lines at the shop floor of plants. This is why HDI regards secondary data and processes only marginally, and to a much greater extent the primary and tertiary contexts. In primary settings, HDI addresses the challenges of fitting "data work" to work, that is of reconciling the articulation of cooperative activities performed by knowledgeable experts with their effort in continuous recording relevant events and information about *what* has been carried out, and often also of *why* and *how*. On the other hand, in the broader ambits of tertiary use, HDI deals with the manifold challenge to design data structures, visualization controls and interaction affordances, for both unintended uses and unanticipated needs, for both the practitioners without particular e-literacy and numeracy, and lay people from the general public.

We would argue that quality of information should not be assessed irrespective of the distinction between primary / secondary / tertiary uses, that is, by adopting the same metrics and methods in a context-independent manner: on the contrary, data used in care processes should be evaluated on the basis of the efficacy they enable appropriate and timely action (fit to use), also on the basis of work conventions and tacit knowledge that are difficult to bring back to the usual dimensions of accuracy, completeness and consistency (Gregory et al., 1995). This is in line with the idea to measure the quality of primary (health) data in terms of appropriateness and as a function of the health outcome they help to achieve. Bad quality should not be related to completeness, timeliness or, even, accuracy, since health practitioners can perfectly make sense of "bad" data and to some extent they even expect them (Bossen et al. 2016); but rather to "information failure" and its impact, that is the number of times data work is either direct or indirect cause of adverse event or near misses (Pipino & Lee, 2011). Despite the authorities that advocated for more research on this concept (Pipino & Lee, 201; Cabitza & Batini 2016), information failure (or error) is still underresearched and a taxonomy of cases to guide analysis and comparison still missing, for instance one that could distinguish between cases in which wrong or late decisions were carried out with accurate and timely data, respectively, and the other way round, a right and timely interpretation has been based on wrong and obsolete data, and so on[5].

Similarly, further research is needed for the evaluation of the quality of tertiary data: this is highly correlated with the quality of the information services conceived for the general public, which should be evaluated in terms of the extent to which data are "informative" and can be appropriated by consumers and appreciated in their lives (i.e., according to its social value), that is if they affect decisions for the better.

This is where HDI can fit in and be a novel way to address long-standing problems of low quality that neither incentivizing structures nor sanctions seem to solve completely. Indeed, HDI covers three phases: design, development and evaluation of the systems by which to extract information and support knowledge in data-intensive application domains. In particular, HDI regards the user-centered elicitation of better requirements of configuration, adaptation and appropriation of big data analytics cockpits and dashboards to optimize usability and the user experience, i.e., efficiency, effectiveness and satisfaction; the application of End-User Development techniques and tools to allow end-users tweak the tools by which data are extracted and visualized; and user-centered methodology for the assessment and continuous improvement of the quality of the interaction of the humans with their data of interest, so as to reduce both information overloading and information funneling/complacency (Parasuraman and Manzey, 2010), and improve awareness. This also includes the exploration of new techniques toward better interactive visualization environments and above all better data-telling, that is the capability to build and share stories that can explain data and facilitate correct interpretations (e.g., in medical domain by adopting a natural frequency approach - Hoffrage et al., 2002) and to allow, instead of curbing, the social exchange within a community of data-users of multiperspective, sound and viable interpretations around the data that are supplied by the computational systems to inform decision making and knowledgeable action.

Summarizing, HDI can contribute in addressing the reconciliation on needs in health data production and use by providing methods and techniques to investigate the following two research strands:

1) The design of interfaces that could promote mutual awareness between data producers and consumers: on the one hand, by increasing commitment and awareness of consequences in the producers of the primary data; and, at the same time, by raising awareness in the secondary consumers of the contextual and social nature of primary data and hence of their limitations. This can be achieved, for example, by endowing the interfaces by which data are collected and presented with specific affordances that adapt to the context according to specific business rules in order to convey the so called "awareness promoting information" (Cabitza and Simone, 2012); this can be done also by means of simple visual clues like text highlighting or side messages, which do not impose any behavior to the data producers but help their interpretation.

2) The design of interactive visual interfaces that could support the understandability of data analytics. The transformation of data into information services does not necessarily require a massive processing of data but rather the application of state-of-the-art human interaction techniques to develop interactive infographics and highly tailorable dashboards that enable user-friendly online analytical processing and hence the transformation, even by end users (Lieberman, Paternò, Klann, & Wulf, 2006), of secondary data into socially valuable information.

## 4. EXAMPLES OF USER STUDIES IN HEALTHCARE HDI

---

[5] Other intersting cases regard the failure to detect relations between either right data or inconsistent ones; the failure to get the right implication from right data or, the other way round, success in detecting right relations and infering right implications from wrong data. It is important to distinguish between all of these cases in understanding the root causes of information failures and understand how to minimize the odds of their occurrence.

| Data Item in the main page of the radiographer | Level of priority |
|---|---|
| Patient ID, Exam Date, Birth Date, Radiographist Name. | HIGHER |
| Level of urgency, Name of the patien, Time of execution, Patient Gender, Responsible Radiologist, Access number, Alert, Body section, DICOM ID, Patient fiscal code, Modality Type, Referring facility. | PROBABLY HIGHER |
| Ordering Radiologist, Exam description, Phone number, Room ID, patient's current address, patient's municipality, how the patient arrived, cost center. | PROBABLY LOWER |
| Patient nationality, Patient height. | LOWER |

Figure 2 The ranking of data item in radiology. Items are enumerated according to their fine-grained ranking within the macro-category of priority level. Levels are established according to statistical testing.

In what follows we report three user studies conceived in a HDI perspective. The first study regards primary data in radiology work, and therefore it regards the identification of what data the practitioners should be required to record for their primary aim, and what data they deem less useful to this aim. The second study regards the value of tertiary data about hospital quality and availability, as this is evaluated by citizens that would need to gain information about the hospitals of their catchment area to choose where to go according to their needs and preferences.

In the first study, we asked 15 expert technicians from different radiology facilities of Northern Italy to indicate the usefulness for their work of 31 information items on an ordinal scale, from 1 (negligible) to 6 (essential). The items regarded either the assessment of patients, their management for the execution of digital imaging examinations, or the preliminary evaluation of radiographic media, and they were collected by considering the Italian guidelines for quality assurance in teleradiology (Working Group for Quality Assurance in Diagnostic and Interventional Radiology, 2010).

The aim of the study was to rank those items from the most useful to the least, so that the interface of their information system could be optimized. Here, the concept of optimization should be interpreted as follows: either in terms of having the radiological information system display and require only those data that the radiographers considered certainly useful for their job; or, at least, in terms of a system that would not urge the users to fill in the low priority fields as mandatory data that block the application unless radiographers insert them. The online questionnaire platform displayed the 31 items in different order to minimize order bias and the task was carried out as part of an assignment of a first Level Master degree class on IT management for radiographers.

We then applied an original ranking method, discussed in (Anonymized) by which we classified each item in one of four levels of priorities (see Figure 2). An informal interpretation of the result could consider the first group (see the green block in Figure 2), i.e., Patient ID, patient birth date, exam date, operator name, radiographer name (if these latter are not the same) as those items that cannot be missing in the radiographer screen, and that indeed need to be emphasized at the interface level (e.g. by being rendered at a proper font size) and double checked by the operators involved to avoid potential mistakes and adverse events. The second group of items (including the name of the patient, the ordering radiologist, the access number) are data of lower priority but still very important, that radiographers should consider mandatory to fill in. Conversely, low priority items like patient nationality and patient height (see the red block in Figure 2) could be considered those that should not be imposed to the data work of radiographers, nor displayed irrespective of the specific examination, as these unnecessary items could contribute to clutter the user interface and hide more relevant information.

In our future work, we will perform the same user study in different groups of users of the radiological information system, like the medical radiologists and the administrative clerks, so as to make the interfaces adaptive according to the role logged in, and have the systems display awareness promoting information (Cabitza & Simone, 2012) about the different perception of utility (if any) by different worker categories. For instance, let us assume that the administrative clerks in a given hospital would consider the indication of the cost center and the billing address of the patient essential data items for their paperwork; in this case, radiographers, who conversely attach a low priority to these items (see Figure 2), could be notified that these data are important for someone else in the same facility (e.g., by means of some iconic indication or by a warning message at the end of the work shift) and be made *aware* (see above) of the data-related needs of other professionals that are connected by the same patient case.

In the second study, we collected 330 complete questionnaires from a population of citizens living in an urban area and aged 18 or older. To this aim, we invited personal friends and acquaintances, all of the colleagues of our university department, and the students of two master classes from the same university. Sample representativeness was achieved by weighting the response set by both age and gender according to the latest national census. The respondents had to evaluate the utility of 11 information items in different scenarios of urgency and serious health conditions. The information items had been selected from those that can be conveniently extracted from some of the most important big open healthcare data sets available in the USA, in Canada and in Italy (taken, respectively, from healthdata.gov, www.cihi.ca and dovesalute.gov.it).

The sample exhibited (statistically) significant appreciation for three information items above all the others: the hospital ranking by number of admissions for the pathology of interest; the ranking by average wait time for the pathology of interest; and the ranking of facilities by their reaching time from one's place. The rankings evaluated the least useful were the rankings produced according to either the perceived level of care quality by all the hospitalized patients, or by the percentage of patients that had to be hospitalized again within six months since the first discharge. While the latter information is quite technical, perhaps too much to be appreciated by a sample of lay people, the former indication came unexpected and quite puzzling.

All of the collected indications are useful to design healthcare portals that show their users a minimal set of items with the highest potential to satisfy the users' needs, on default of other information, leaving the other items accessible on demand. Interestingly, we detected significant differences in the perception of information usefulness between different user profiles. For instance, male respondents attached more importance than women to the ranking based on georeferenced information. Not surprisingly, expert users considered the ranking by number of admissions for the pathology of interest more important than people self-declaring non-expert. We also probed the sample preferences for different ways to display the same information, either by simple and essential tables or by richer and more appealing infographics. Also in this case we detected a statistically significant difference between females and males, as the former ones found infographics to be clearer and more intuitive than men; and between young respondents and the elderly as the former ones, quit unexpectedly, found the tabular form more clear than the latter ones. These differences suggest that the usability of a healthcare portal can be maximized even considering basic profiling information and user satisfaction can be improved by asking the user only a few data, like gender and age.

In the third case we involved 29 General Practitioners: 5 out of 10 have been family doctors for more than 30 years, 9 out of 10 for more than 20 years; no respondent had less than 10 years of work experience so that we can consider the sample composed of experts in the Family Medicine domain. The majority of the respondents (65%) claimed to be competent in Information Technology (IT) but neither experts nor enthusiasts. These latter ones were, respectively, approximately one third and one fifth of the sample.

The respondents were asked to choose the three most useful datasets to be visualized on a personal dashboard for their daily practice by selecting them out of a list of 13 data sets. These had been selected for their popularity in terms of number of downloads from a comprehensive collection of Italian health open datasets[6]. The sets chosen more frequently by the respondents were:
1. Regional distribution of the percentage of elderly people involved in the program of Integrated Home Care (in Italian: Assistenza Domiciliare Integrata).
2. Number (and details) of the Rehabilitation facilities for the elderly (in Italian: Istituti di Riabilitazione Extraospedaliera per Anziani).
3. Inpatient care average length of stay.

These datasets were selected by approximately one third of the sample (38%, 33% and 30% respectively). The other datasets collected much fewer preferences, with the "number of cases affected by pneumococcal disease by year" that was not chosen by any respondent. However, the datasets that the respondents found to be more useful for general practice to be georeferenced on an interactive map were others, namely:
1. Number of registers on exposure to carcinogens (.5 vs. .95, p=.000).
2. Hospitalization rates, by regime, patient genre and region (.26 vs. .74, p=.035).

As a visualization aid to the comprehension of patient data and their correlation on multiple dimensions, parallel coordinates, like those depicted in Figure 3 and 4 were found to be highly informative: the responses were positive for the majority (.64 vs. .46, p=.286). Furthermore, almost every GP involved in the user study claimed that the interactive version of the diagram was much more informative than the static view (.96 vs. .04, p=.000, 60% "much more informative").

---
[6] http://www.datiopen.it/en

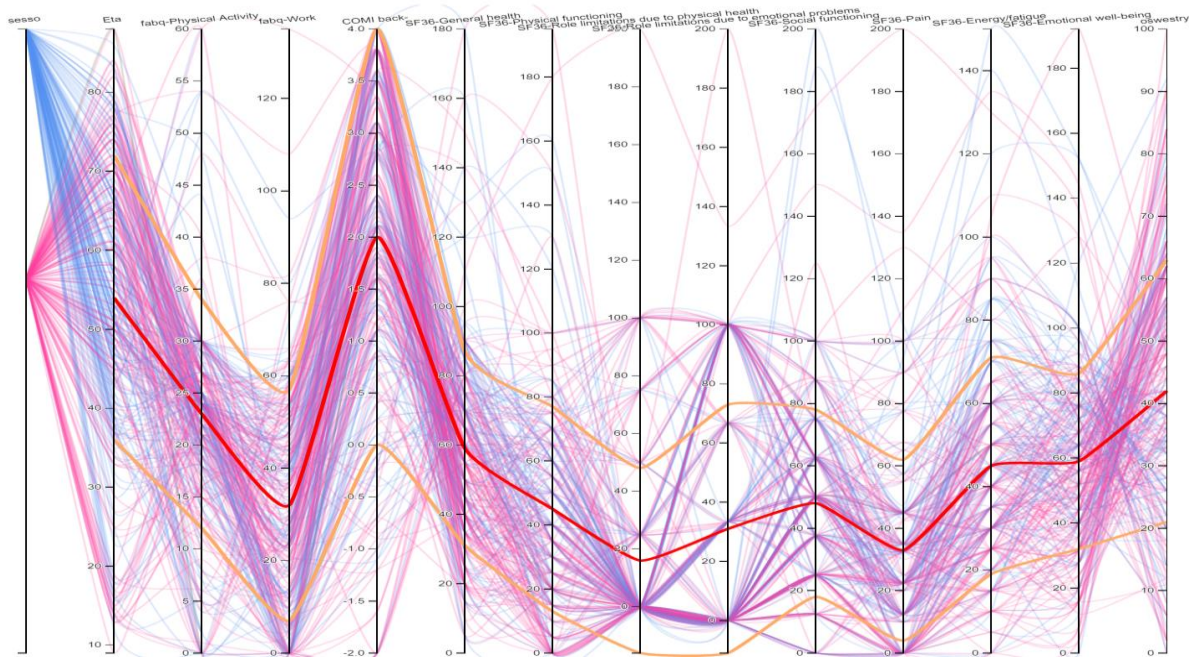

Figure 3. The parallel coordinate diagram shown to the user panel. The thick red and orange lines indicate the average values and standard deviations respectively. Blue lines represent single male patients. Pink lines female patients.

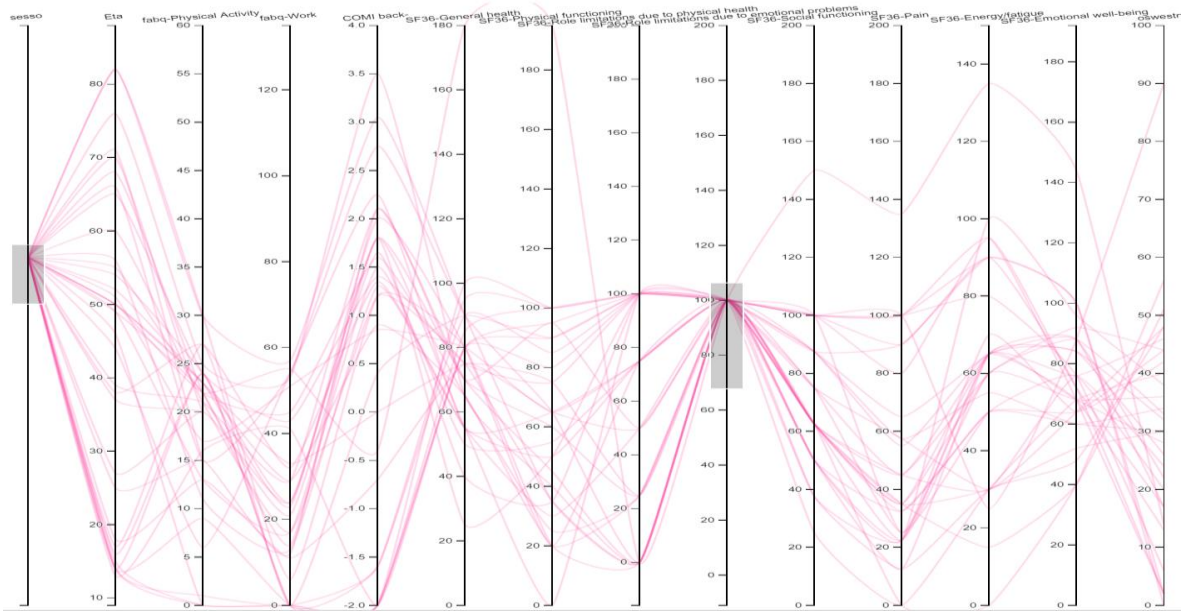

Figure 4. The parallel coordinate diagram shown above after interacting with it. In particular two filters were created: one on the first axis, to select only female patients; another to select only patients whose responses to a SF36 form scale were between 70 and 110.

## 5. CONCLUSIONS

In this paper we have proposed a triparted perspective to health data by which to distinguish between primary data, i.e., the content of health records that clinicians and nurse produce and consume in their decisions and treatments; secondary data, i.e., the content of structured data bases, data warehouses and the reports that automated procedures produce and display on interactive data visualization tools (or dashboards) to support the tasks of clerical and managerial roles (including policy makers); and tertiary data, i.e., particular content that has been selected and optimized in terms of clarity, understandability and appeal, in one word usability, to be offered online for the general public and hence for unanticipated needs and aims. The distinction has been proposed to separate concerns and differentiate agendas. The one-size-fits-all approach to data quality improvement simply does not work (Pipino et al. 2002), nor solve the still present problems of low use of and low satisfaction for the Electronic Health Record (Cabitza et al. 2015a). Moreover, such a simple separation of concerns also allows to emphasize the concept of "use and reuse" of health data (Berg and Goorman, 1999), which is important to keep in mind when setting quality requirements and imposing ideal data model to health practitioners and their patients.

Human Data Interaction (HDI), as a research approach, can provide alternative methods to address these challenges, which are becoming more and more urgent especially in our current age of increasing opportunities and threats regarding the datification of medicine and the consequent production of health big data (Murdoch & Detsky 2013). In particular, HDI approaches would focus on evaluating how ICT supports data (producing) work by relieving doctors and nurses from form-filling, coding and other administrative tasks; and on evaluating the usability and the user experience related to interacting with tertiary (health) data, to understand the actual value these convey in decision making and information-intensive tasks.

## REFERENCES

Abiteboul, S. et al, 2000. *Data on the Web: From Relations to Semistructured Data and XML*. Morgan Kaufmann Publishers, San Francisco, USA.

Abdelhak, M., Grostick, S., & Hanken, M. A. (Eds.). (2012). Health information (4th ed). Elsevier.

Berg, M. (1999). Accumulating and coordinating: occasions for information technologies in medical work. Computer Supported Cooperative Work (CSCW), 8(4), 373-401.

Berg, M., & Goorman, E. (1999). The Contextual Nature of Medical Information. IJMI, 56, 51–60.

Berg, M., & Toussaint, P. (2002). The mantra of modeling and the forgotten powers of paper: A sociotechnical view on the development of process-oriented ICT in health care. JMI, 69(2-3), 223–34.

Bossen, C., et al. (2016) Data-work in Healthcare: The New Work Ecologies of Healthcare Infrastructures. ACM International Conference on Computer-Supported Cooperative Work and Social Computing, CSCW 2016.

Cabitza, F., & Simone, C. (2012). Affording Mechanisms: An Integrated View of Coordination and Knowledge Management. CSCW, 21(2), 227–260.

Cabitza, F., Simone, C., & De Michelis, G. (2015a). User-driven prioritization of features for a prospective InterPersonal Health Record: perceptions from the Italian context. Comput Biol Med, 59, 202-210.

Cabitza, F., Locoro, A., & Batini, C. (2015b). A User Study to Assess the Situated Social Value of Open Data in Healthcare. Procedia Computer Science, 64, 306-313.

Cabitza, F., & Batini, C. (2016). Information Quality in Healthcare. In Data and Information Quality, 421–438. Springer.

Crabtree, A., & Mortier, R. (2015). Human data interaction: Historical lessons from social studies and CSCW. In ECSCW 2015. Springer.

Fitzpatrick, G. (2004). Integrated care and the working record. Health Informatics Journal, 10(4), 291-302.

Fitzpatrick, G., & Ellingsen, G. (2013). A review of 25 years of CSCW research in healthcare: contributions, challenges and future agendas. CSCW, 22(4-6), 609-665.

Gitelman, L. (2013). Raw data is an oxymoron. MIT Press.

Gregory, J., Mattison, J. E., & Linde, C. (1995). Naming notes: transitions from free text to structured entry. MIM, 34(1-2), 57–67.

Hoffrage, U., Gigerenzer, G., Krauss, S., & Martignon, L. (2002). Representation facilitates reasoning: What natural frequencies are and what they are not. Cognition, 84(3), 343-352.

Liaw, S. T., Rahimi, A., Ray, P., Taggart, J., Dennis, S., de Lusignan, S., ... & Talaei-Khoei, A. (2013). Towards an ontology for data quality in integrated chronic disease management: a realist review of the literature. IJMI, 82(1), 10-24.

Lieberman, H., Paternò, F., Klann, M., & Wulf, V. (2006). End-User Development: An Emerging Paradigm. In EUD, 9, 1–8.

Locoro, A. (2016) A Map is worth a Thousand Data: Requirements in Tertiary Human-Data Interaction to Foster Participation. CEUR Workshop Procs of CoPDA 2015.

Murdoch, T. B., & Detsky, A. S. (2013). The inevitable application of big data to health care. Jama, 309(13), 1351-1352

Parasuraman, R., & Manzey, D. H. (2010). Complacency and bias in human use of automation: An attentional integration. Human Factors: Hum. Factors, 52(3), 381–410.

Pipino, L., Lee, Y. W., & Wang, R. Y. (2002). Data quality assessment. Commun ACM, 45(4), 211-218.

Pipino, L., & Lee, Y. W. (2011). Medical Errors and Information Quality: A Review and Research Agenda. In AMCIS.

Roberts, A., Gaizauskas, R., Hepple, M., & Guo, Y. (2008). Mining clinical relationships from patient narratives. BMC Bioinformatics, 9(11), S3.

Sauleau, E. A., Paumier, J.-P., & Buemi, A. (2005). Medical record linkage in health information systems by approximate string matching and clustering. BMC Medical Informatics and Decision Making, 5(1), 32.

Shankaranarayan, G., Ziad, M., & Wang, R. Y. (2003). Managing data quality in dynamic decision environments: An information product approach. JDM, 14(4), 14–32.

Shortliffe, E.H., Perreault, L.E., Wiederhold, G., and Fagan, L.M. (eds.) (1990). Medical Informatics: Computer Applications in Health Care and Biomedicine. Addison-Wesley.

Smith, P. C., Araya-Guerra, R., Bublitz, C., Parnes, B., Dickinson, L. M., Van Vorst, R., Pace, W. D. (2005). Missing clinical information during primary care visits. Jama, 293(5), 565–571.

Swinglehurst, D., Greenhalgh, T., & Roberts, C. (2012). Computer templates in chronic disease management: ethnographic case study in general practice. BMJ Open, 2(6)

Timmermans, S., & Berg, M. (2003). The gold standard: the challenge of evidence-based medicine and standardization in health care. Temple University Press.

Wang, R. Y., & Strong, D. M. (1996). Beyond accuracy: What data quality means to data consumers. JMIS, 5-33.

Wang, R. Y. (1998). A product perspective on total data quality management. Commun ACM, 41(2), 58-65.